# The Scientific Impact of Einstein's visit to Argentina, in 1925

Alejandro Gangui (1) and Eduardo L. Ortiz (2)

(1) Instituto de Astronomía y Física del Espacio, Conicet and University of Buenos Aires,
    C.C. 67, Suc. 28, 1428 Buenos Aires, Argentina

(2) Imperial College, London,
    South Kensington campus, London SW7 2AZ, England

The arrival of Albert Einstein in Argentina in 1925 had an impact, equally relevant, on the scientific community and on the general public. In this paper we discuss that visit from three different perspectives. Firstly, we consider the conditions that allowed for such visit to be possible. Then we focus on the institutional actors that facilitated it, as well as on the expertise and written references on topics related to relativity theory circulating at the time in the local community. In the last section we consider the implications of that visit for the local scientific environment.

## Introduction: Scientific visitors in Argentina

In parallel with Argentineans who moved to Europe to study in its centers of learning, the history of that country records, over a very long period which goes back to Colonial times, the presence of a long sequence of scientific visitors who, for the most part, were naturalists and cosmographers. When the country became independent the string of visitors was not interrupted, rather, it was firmly stimulated, particularly, if those visits could contribute to its scientific or technical advance. From the area of culture, among the most frequently stated argument for independence was the difficulty experienced to attract science and technology teachers, or to be able to send local students to train in these areas in Europe. These barriers, placed by the Crown, prevented the recruitment of science and mathematics teachers who could boost the development of industries, a proper division of land, and training on astronomical navigation.

     One of the first actions of Bernardino Rivadavia (1780-1845), the first Argentine envoy to Europe, who later was the first President of Argentina, was to invite leading scientists to teach in Argentina's capital. He managed to attract from Europe a well known mathematician, the Spaniard José María Lanz (1764-1839), and an equally distinguished physicist, the Italian Ottaviano-Fabrizio Mossotti (1791-1863). Later many other eminent scientists reached the country, among them stand prominently Aimé Bonpland (1773-1858) and Charles Darwin (1809-1882); the first, a renowned world scientist that had accompanied Alexander von Humboldt (1769-1859) in his travels through the American continent.

     After a difficult period, which covered the decades of 1830 to 1850, the number of scientific visitors, and the impact of their presence in the country, multiplied. Scientific work at the Public Museum and at the University of Buenos Aires began to revive and, later, reached also the old University in the city of Cordoba, where a new scientific department was created in the 1870s.

     Meanwhile, advances in security and in the speed of sailing, particularly with the spread of the clippers and other fast vessels, made it possible to receive visits of European personalities in Argentina for considerably shorter periods than in the past: visits now could be worthwhile for periods of just a few months, or even weeks, not years as before. A very interesting scientific short visit of that period was that of the English engineer John Bateman (1810-1889), who visited

Buenos Aires for less than two months in 1870-71 to advise on the design of the city's port and its water supply system. In the field of science and technology this visit was the herald of a new type of expertise exchange, brief but able to produce a significant impact.

Almost simultaneously, in 1875 and perhaps not unrelated to the previous trip of Bateman, Argentine naturalist Eduardo L. Holmberg (1852-1937) conceived in his novel Two Parties at War ("Dos Partidos en Lucha") (Holmberg, 1875), a scientific fantasy based on the idea of Darwin travelling to Buenos Aires aboard a fast, modern steamship, the *Hound*, capable of reaching 30 miles per hour to participate in a debate on evolution.

The fact is that brief visits to Argentina by European specialists, for specific scientific purposes, became more frequent only a couple of decades later than imagined by Holmberg, when replacing of sail by steam allowed for the opening of regular, reliable and cheaper services between Europe and South America. This began to happen when the achievements of science and technology at home created a sufficiently large local audience to justify these trips.

The Independence centenary celebrations of 1910 offered an opportunity for the dreams of Holmberg to begin to be realized. Among the members of foreign delegations who visited Argentina for a short time were two eminent scientists: one of them, accompanying the Spanish delegation, was Leonardo Torres Quevedo (1852-1936), whose invention of an electro-mechanical chess-playing machine contributed to opening the modern world of automation and computing; he lectured on the workings of one of his automatic electro-mechanical calculating devises. The other, accompanying the Italian delegation, was the mathematician and theoretical physicist Vito Volterra (1860-1940). He was one of the first relativists whose work we shall mention later. In a significant lecture (Volterra, 1910), he introduced the modern ideas linking space, time and matter, which he himself was helping to develop, and in it mentioned the name of the young Albert Einstein.

The Independence centenary and the steam engine also contributed to make possible a series of short visits by Spanish intellectuals, many of them trained rigorously in Germany. They were sponsored by the Spanish Cultural Institution of Buenos Aires (Ortiz, 1988), an institution supported by local Spanish merchants. In series of lectures that typically lasted for a few months (usually covering the recess of activities in Europe) these intellectuals began to renew Argentina's culture in areas as diverse as science, philosophy, sociology and linguistics, and they did so with a breadth and depth that had never been seen before there. Among those visitors were the philosopher José Ortega y Gasset (1883-1955), the mathematician Julio Rey Pastor (1888-1962) and the physicist Blas Cabrera (1878-1945). These three were also major players in the story that we will cover in this work.

After the end of the World War I the Spanish visits continued, and in the 1920s were emulated by visits of French, Italian and German intellectuals. Among them: Albert Einstein (1879-1955), Paul Langevin (1872-1946), and Émile Borel (1871-1956). The visit of the first of these scientists is the subject of this paper.

These leading scientific visitors left also a mark in the daily press and in magazines of the time. These publications recorded their thoughts and an image of their human and spiritual personality. In their eyes, whether they were musicians, writers, politicians or scientists, they all had singular personalities together with very unusual and exceptional abilities. The artists were shown as able to convey complex feelings; politicians, usually extraordinary speakers, able to offer solutions to all the big problems of the World, while scientists gave local people reassurance that progress was, really, unlimited.

Sometimes it was difficult to communicate with them, or understand the meaning of their message. However, for those who were already famous there were, at the offices of the major local newspapers, clippings from the foreign press that facilitated the reception of their message and allowed local journalists to repeat about them, with a certain margin of safety, what had already been said in their previous visits to some of the great capitals of Europe.

In the case of Einstein the message was complex, no only because of the unusual formal and intellectual complexity of his ideas but also because his human image did not fit into a firmly established picture of the sage either. That image, consolidated by a legion of previous visitors, did not include such and open personality with a tendency to use irony, even with the press interviews. This particular visitor forced a revision of both, the image of spacetime and that of the wise man.

This fascinating process, of which we shall try to give a sketch here, took place in Buenos Aires, and to a lesser extent in the cities of La Plata and Córdoba, in the early months of 1925. It left a mark that is still debated. Further details on this visit can be found in our previous research on this subject (Ortiz, 1995; Ortiz and Otero, 2001; Gangui and Ortiz, 2005, 2008, 2009, 2011) and Gangui, 2007).

## Why did Albert Einstein go to Argentina in 1925?

The year 1919 is remembered as a turning point in the public figure of Einstein. On May 29 of that year a total solar eclipse allowed astronomers to test one of the most important predictions of his theory of general relativity: the deflection of light in a gravitational field. After the theory was proposed in late 1915, and after several failed astronomical attempts to test it, in which astronomers from Argentina were not absent (Gangui and Ortiz, 2009), in 1917 Arthur S. Eddington and the Astronomer Royal, Frank Watson Dyson, suggested sending an expedition to observe the total eclipse of 1919 to try to verify the above mentioned astronomical prediction of Einstein's theory. Two expeditions were sent to different areas where the eclipse was visible in its totality: the first to Sobral, in the state of Ceará, in North Brazil, was led by Andrew Crommelin and Charles Davidson, and the second to Principe Island, in West Africa, off the coast of Spanish Guinea, was led by Edwin Cottingham and Eddington. The latter was then a prestigious figure who knew Einstein's theory well.

The result of the delicate astronomical observations, as expected by the expeditionary teams, suggested that in the vicinity of the solar disk the positions of the stars selected for observation were slightly removed from the center of said disc. In addition, the tiny displacement observed was compatible with the predictions of relativity theory. The deviation showed space, and therefore also the paths of the light beams of the stars, "bend" in the presence of the Sun.

In England, the announcement of the results was made at a special ceremony organized jointly by two prestigious institutions: the Royal Society and the Royal Astronomical Society; it took place in London on November 6, 1919. The significance of this meeting, which took place within a year of the end of the First World War, is huge: a group of British astronomers led by pacifist Quaker Eddington presented evidence supporting the extraordinary predictions of a new theory proposed by another pacifist -but German. In addition, it corrected the old mechanical laws formulated by Newton, which had reigned supreme for several centuries.

## The beginnings of relativity in Argentina

Towards the end of the nineteenth century, well before Einstein's work, studies on spacetime and radiation regularly appeared in Europe and, later, began to be transmitted elsewhere. Among the early published work in that field in Argentina were a couple of papers of the Argentine mathematician Valentin Balbin (1851-1901), who had received training in England, and of the engineer and physicist Jorge Duclout, a French Alsatian who had graduated from the Zürich ETH. Both were then professors at the University of Buenos Aires and ran a graduate seminar on mathematical-physics at the local Scientific Society (*Sociedad Científica Argentina*), an institution founded in 1872 in Buenos Aires which was, for a considerable period of time, the champion of science in that city.

In the first decade of the twentieth century there appeared in Buenos Aires a few notes on matter and radiation, and also on the new (special) theory of relativity. These works seemed to offer a unified view of matter and energy and introduced new concepts about the structure of physical space. Like those of Balbín and Duclout, they were technical and informative papers published in locally prestigious scientific journals sponsored by the Scientific Society, the university or the engineering students' union. Authors included both foreigners and local people and in general were accredited academics (Galles, 1982).

Among the papers on issues related to relativity that appeared in Argentina before Einstein's visit that of the Italian mathematician Ugo Broggi (1880-1965), published in 1909, stands out. In it he accurately discussed contemporary ideas on matter, radiation and time. Broggi had received a doctorate at Göttingen University working under David Hilbert on a topic of probability theory; he later moved into the mathematics of finance. By the time he published that paper he was a professor of mathematics at the University of La Plata, a new university, created in 1905 as part of a drive to introduce modern science, particularly experimental science, in Argentine universities.

A year later another Italian, mathematician and physicist Vito Volterra, already mentioned in connection with relativity theory, visited Argentina as a member of the Italian delegation to the celebrations of the first centenary of Argentina's independence. At the time he was also a Senator in his country's parliament; while in Buenos Aires, Volterra was invited by the Argentine Scientific Society to lecture on relativity. In his lecture (Volterra, 1910) he dealt with spacetime and matter, and referred specifically to the recent work of Einstein.

Between 1910 and 1915 the French mathematician and physicist Camilo Meyer (1854-1918), who had been a fellow student of Henri Poincaré and kept in touch with him while in Argentina, offered, for several years, an unofficial course on mathematical physics at the University of Buenos Aires (UBA). In the last of them, which fortunately has been preserved (Meyer, 1915), he gave an account of recent advances in physics, including the "theory of quanta". In that final course he made explicit references to Einstein's work but, due to the specific topics he developed, mainly of interest to chemical physicists, he did not mention Einstein's work on the theory of relativity.

By the end of the first decade of 1900, the prestigious American astronomer Charles Dillon Perrine (1867-1951), formerly of Lick Observatory, California, was appointed director of the National Astronomical Observatory, at the Argentine city of Cordoba. The latter observatory, created in the early 1870s, was then receiving considerable support as part of the already mentioned drive to promote the development of experimental science. Perrine renewed the optical workshops and observation facilities of the institution and, in 1912, answering a request from colleagues in Germany associated with Einstein, he began a series of observations, in successive solar eclipses, aiming at detecting possible (relativistic) deflections of light. His observations began in Brazil in 1912, continued in Russia in 1914 and ended in Venezuela; however, due to unfavorable weather conditions Perrine and colleagues could not provide a conclusive answer (Gangui and Ortiz, 2009). Sadly, due to financial difficulties affecting Argentina after the end of the First World War, he could not find support to take part in the solar eclipse of 1919, in nearby Brazil, when the shift suggested by the general theory of relativity was finally detected.

Another speaker in favor of relativity in Argentina was the physicist Jakob Laub (1884-1962), hired by the University of La Plata as professor of physics in the early 1910s as part of the same movement for experimental science. Before moving to Argentina Laub had personal contact with Einstein (Seelig, 1954); he collaborated with the latter in early research on relativity theory (Pyenson, 1985). Once in Argentina, Laub published a series of papers on that theory in the journal of the Argentine Scientific Society. His was the work of a scientist who knew the theory of relativity and the ideas of Einstein in detail (Gangui and Ortiz, 2011). However, as the decade progressed and the public interest in relativity grew, his papers moved from specifically science journals to qualified mainstream media, for example, *Revista de Filosofía* (The Philosophy Journal), then led by the prestigious medical doctor and positivist philosopher José Ingenieros (1877-1925). That journal, a landmark of culture in Argentina, included contributions targeted to the interests of well educated readers, but had a much broader cultural focus than just theoretical physics.

As another important precedent for the introduction of relativity theory at a philosophical level, we should also mention the academic conferences on new trends in German philosophy given by the celebrated Spanish philosopher José Ortega y Gasset at the Humanities Faculty in Buenos Aires (Ortiz, 2011b); they had a considerable impact on the Argentine cultural world and helped move the philosophical debate on the impact of relativity theory to a new, much higher level.

The lectures given by the Spanish physicist Blas Cabrera during his visit to Argentina, in 1920 (Ortiz and Rubinstein, 2009), had also a considerable impact. Taking advantage of the difference in seasons, Laub spent several European winters working in Cabrera's laboratory in Madrid.

The work of the Spanish Jesuit astronomer José Ubach (1871-1935) on the results of the astronomical expedition of 1919 should also be mentioned; it informs us about the contemporary position of the Catholic Church on that subject. Ubach had been trained at the Ebro Observatory and since the 1910s was a science teacher at the prestigious *Colegio del Salvador*, Buenos Aires; he had also access to astronomical observation instruments and observed several eclipses of the Sun from Argentina.

A few years later two young Argentine scientists trained in physics in Germany, José B. Collo and Teófilo Isnardi, published a descriptive study of relativity, in which Einstein's theory is discussed from a more technical point of view. On account of its quality, this piece is indicative of both an important intellectual shift in the direction of the *theoretical* sciences, that was taking place in Argentina precisely in these years, and also of certain maturity in the discussion of relativity theory. Their work included a third part, written by Félix Aguilar (1884-1943), an Argentine astronomer who had also been trained in Europe; he discussed the astronomical implications of relativity. This key work has been discussed in some detail in (Gangui and Ortiz, 2009; 2011).

## The public figure of Einstein in Argentina

From the year of "Einstein's eclipse", news and articles on various aspects of the theory of relativity began to appear regularly in the cultural sections of the leading local press. We have already suggested that there was an interest in scientific literature among Argentine educated readers and also amongst authors (Gangui and Ortiz, 2011). The leading postmodernist poet Leopoldo Lugones (1874-1938) was, no doubt, primary responsible for the continued presence of these topics in the local media; his interest in science and in Einstein's relativity in particular, is well documented (e.g., Asúa and Hurtado de Mendoza, 2006). Moreover, that prominent intellectual had considerable influence on the editors of several large circulation newspapers in the country, many of whom were also distinguished poets and writers.

Although Lugones' scientific readings were wide, this did not make him an expert, let alone a reference on topics related to contemporary theoretical physics. However, he was a respected intellectual and his judgment on almost any topic, including science, had a considerable weight. In 1920 he was invited by the *Centro de Estudiantes de Ingeniería* (the University Engineering Student's Union, which also covered science studies), to give a lecture on the implications of the theory of relativity. Let us note, in passing, that this students' union played a prominent role in the cultural landscape of the first half of twentieth century Argentina.

On the basis of that lecture, a year later Lugones published a small, very well written book with the suggestive title "The size of the Space" (*El tamaño del espacio*) (Lugones, 1921). Although the book was not free of misinterpretations and of somewhat bizarre and even esoteric passages and quotations, as often in Lugones' fantasy work, the book served to place relativity in a favorable position vis-à-vis the audience for which it was intended. The mere fact that an intellectual of the caliber of Lugones was interested in a theory so far removed from everyday life, was much more important for the reception of the theory than accuracy in the description of the physics to which the poet could aspire (Ortiz, 1995). His work was followed by others on relativistic issues, written by local authors or translated from foreign languages.

In the leading local newspaper *La Nación* of June 23, 1921 a Spanish reporter indicated that Einstein was aware of Lugones' work and even thanked him for the kind references he made to him in *El tamaño del espacio*. That year, and in the following ones, the main Buenos Aires newspapers regularly included news about the many tributes the German scholar received in different countries, particularly in France, which at that time was Argentina's preferred compass reference.

After the success of Lugones' lecture, the Engineering Student's Union invited the already mentioned Duclout, who had also been a professor of elasticity theory, to read a paper on matter-energy and relativity, published in 1920 (Duclout, 1920).

Duclout, like his friend Lugones earlier, together with the mathematician Julio Rey Pastor, played an important role in making Einstein's visit to Argentina possible (Ortiz, 2011a). Close to Einstein's visit, Rey Pastor published in local newspapers, particularly in *La Nación*, useful notes on topics related to relativity. He had met Einstein personally in Berlin and played an important role in the invitation for Einstein to visit Spain, which he did in 1923.

The public interest in the issues that made Einstein a scientific celebrity had an impact in the local world of university students. The Engineering Student Union, apparently at the suggestion of Rey Pastor, wrote to Einstein on April 5, 1922 asking for permission to translate and publish in Spanish his highly technical work "Die Grundlage der allgemeinen Relativitätstheorie", originally published in the German journal *Annalen der Physik* in 1916; they indicated that amongst some of its members there was an interest in that topic. Einstein responded quite soon, by May 31, accepting the proposal and suggesting that the publication be made in book form; as royalties he asked for 20% of the profits.

## Local reactions against threats to Einstein

By 1922 the political and social situation in post-war Germany was difficult; newspapers in Argentina kept their readers informed of the political instability of the country and, towards the end of June, gave news of the assassination of Walther Rathenau (1867-1922), Foreign Minister of the Weimar Republic. *La Nación* indicated also that Einstein's life was in danger, a possibility that was repeated in the following months. Towards the end of 1923 Einstein had, in fact, to leave his country and travel temporarily to the Netherlands as a consequence of threats he received.

Aware of the situation, in August 9, 1922 Lugones published an article in *La Nación* suggesting that the sage be invited to Argentina; local friends of science should contribute with funds to cover the expenses to create a independent university chair specially organized for him. Shortly after this note, two student groups in Buenos Aires: the university Engineering Students' Union and its sister at the Secondary School Teachers' Training National Institute (*Instituto Nacional del Profesorado Secundario*), publicly announced their support for Lugones's idea to offer Einstein a safe place to live and work until the situation in Germany became more stable. Two renowned scientists, mentioned before, were unofficial councilors of these student's unions: Rey Pastor, to the first, and Laub to the second; at the time he ran its physics department. Duclout, Lugones' personal friend and then closely connected with university councils, was also a key element in this movement of opinion which, as we have just indicated, combined young university students and a group of leading intellectuals.

On August 22 of that year Duclout submitted a proposal to the Sciences Faculty Board, in Buenos Aires, to grant Einstein an honorary doctorate degree in the mathematical and physics sciences for his work on the theory of relativity. The proposal was endorsed unanimously and in just a few days, was also accepted by the University Council. The literature professor Mauricio Nirenstein, who played a significant role during the visit of the German scholar in Buenos Aires, was at that time the University's secretary.

A few months later, on October 30 Duclout, supported by other scholars, submitted to the Council of Buenos Aires University a formal proposal suggesting that Einstein be invited to give a series of lectures on the theory of relativity. The timing could not have been better, as in less than two weeks local newspapers announced Einstein had been awarded the Nobel Prize for Physics for the year 1921, which was left vacant the previous year. Let us recall that the German scientist did not receive the award for his work on relativity, which at the time was not universally accepted even after astronomical evidence, but for his explanation of the photoelectric effect.

At the time Einstein was embarking on long trips abroad: to England and the United States in 1921, to France and Japan in the following year and Spain, Japan and Palestine in 1922-23. Argentine newspapers continued to follow, in detail, the movements of the wise man.

A year later, on December 21, 1923, in a extraordinary session, the Council of Buenos Aires University discussed, among other topics, Einstein's visit to Argentina, considering as well a note from the *Asociación Hebraica Argentina* (Argentine Hebraic Association) indicating they had already established contacts with the father of relativity and had offered him the sum of US$ 4,000 and two ship tickets for a visit to Argentina. The Association also noted that the scientist was interested in undertaking the trip only if the invitation was extended by an academic institution; knowing the interest already shown by the University, the Hebraic Association considered it appropriate that the former be in charge of the formal invitation.

To help with the logistics, the Hebraic Association would contribute 4,660 pesos (the equivalent of about US$ 1,500 at the time) to meet part of the fees. The Council of the University agreed to contribute US$ 2,500 to provide for Einstein's visit, an amount that, together with the smaller contribution of the Hebrew Association, totaled the required US$ 4,000. This sum had been considered appropriate to enable his visit to the country. In Argentina, it amounted, approximately, to a full year contract of a teacher of the highest level, say, of the stature of Rey Pastor. The University also agreed to cover half the cost of the two ship tickets (for Einstein and, eventually, his wife), in case the negotiations with the Argentine government to raise funds from it were not successful. Months later, in January 1924, the Argentine-Germanic Cultural Institution offered also a contribution (approximately 1,500 pesos) to help finance the visit.

From that time onwards, the Council of the University received frequent news on the state of the negotiations with the prospective visitor. In May 1924, it was informed of Einstein's correspondence with the representative of the Ministry of Foreign Affairs of Argentina in Berlin and with the Hebrew Association. Finally, the visit was set for the end of March next year, that is, towards the beginning of the 1925 academic year.

It is interesting to note that, in order to explain what the theory of relativity really was about, Argentina, then an exceedingly wealthy country, sought, precisely, the word of the creator of that theory. Just a few years later another eminent physicist, the French relativity expert Paul Langevin, travelled to Buenos Aires sponsored by the *Institut Français de l'Université de Paris à Buenos Aires*. His visit was made at far smaller cost for the university. In addition, Langevin rejected a payment the university offered him for a set of advanced lectures.

Einstein's visit to Argentina provides also important clues both to the history of the development of the Jewish community in that country and also to its relations with contemporary intellectual movements. We will not discuss these matters here, which the interested reader could find in (Ortiz, 1995) with original material from local and foreign archives.

## The impact of Einstein's visit

In addition to the significant changes that had taken place in the international scene in the years immediately before Einstein's visit, there were also significant changes inside Argentina, where the universities were still feeling the renovating impact of a University Reform Movement, started by university students there in 1918, which also had a considerable cultural impact on other countries of Latin America. On account of these circumstances, the general public perceived the world of Einstein's ideas as part of an important intellectual revolution characterized by a fundamental change in the ways in which reality was understood. Despite the fact that the theory carrying his name was based on almost imperceptible facts, which then seemed to have no connection whatsoever with ordinary life, he was listened to with attention, with respect and even with guarded admiration.

It could not be said that his ideas were generally understood with any fidelity, or in the same way in which they were perceived then in some of the leading countries of Europe. Writing about his perplexing theory, local newspaper and magazine reporters could not go beyond exposing their own limitations, while philosophers attempted to express more coherent and sophisticated thoughts, not always with success. However, both groups perceived clearly that Einstein's ideas seemed to show that it was necessary to think again, from scratch, about the nature of scientific knowledge and physical reality. Amongst Argentine scientists only a few could follow the complex formulism and the physical ideas Einstein used in his theory and in his lectures. However, in academic circles some clearly realized they were taking part in a unique event, and showed determination to make an effort to try to understand. Some magazines and newspapers made his lectures more widely available by publishing them in full versions from notes taken by young, German-trained physicists.

The visit had other consequences; one of them, affecting both the university community and political circles in charge of designing cultural policies, was a clear understanding that research in the *theoretical* sciences, mathematics, theoretical physics, was a cultural component which had been neglected in the past.

We have already pointed out in previous work (Ortiz, 2011b; Gangui and Ortiz, 2011) that Einstein visited Argentina at a very special moment in time, when important conceptual changes were taking place at the highest levels of Argentina culture. It began to separate itself from the previously dominant conceptions of positivist philosophy, or rather of a vernacular version of it, which had been dominant for at least twenty years. In the fields of science that particular branch of philosophy had helped promoting the creation of large *experimental* research centres. This was the case in the leading countries of Europe and also in Argentina where, in the first decade of the 20th century and in the specific field of the exact sciences, that trend is reflected in the creation of a new and modern Physics Institute at the University of La Plata and in the renovation of the National Astronomical Observatory, at Córdoba, with new instrumentation and workshops; well known scientists were imported to accelerate the changes.

Einstein's visit contributed to reaffirm a perception of the *theoretical sciences* as the engine for fundamental changes in relation to our conception of the world and nature; however, there are clear indications that some of these perceptions already existed *before* the visit and indeed helped to motivate the efforts made to precipitate it.

In Buenos Aires, a series of lectures given at the Humanities Faculty by the German trained Spanish philosopher José Ortega y Gasset were amongst the first efforts to break the dominance of Positivism. Since the Reform Movement on 1918, a sustained effort had been made in support of modernizing mathematics, physics and philosophy training at university level. We have already remarked the insistence of philosopher Coriolano Alberini on setting a clear definition vis-à-vis Positivism from Einstein (Gangui and Ortiz, 2011).

In 1917, over several months, the Spanish pure mathematician Julio Rey Pastor, recently trained in Göttingen and Berlin, gave a series of lectures on advanced pure mathematics at the University of Buenos Aires. Although centred at the Science Faculty, Rey Pastor was also invited to lecture at the Humanities Faculty. His lectures there were based on a recent book of his (Rey Pastor, 1989) in which he discussed some of the leading questions of contemporary mathematics: the foundations of geometry, algebraic structures, the problems of infinity. This set of lectures attracted a large and exclusive audience which included some of the leading intellectuals of the time showing, again, the existence of a new interest in the recent advances of the theoretical sciences, which can be detected through most of the decade of the 1920s. At the end of his series of lectures and seminars, Rey Pastor was invited to join the University of Buenos Aires on a permanent basis, which he did in 1922.

Rey Pastor's presence in Buenos Aires precipitated the development of a mathematics department in the university, where pure mathematics was emphasized. Only a few years later, soon after Einstein's visit, a new physics department was created there. In these new environments modern topics of pure mathematics and theoretical physics began to be discussed and developed as independent areas of research. Their possible impact on the advancement of technology was, no doubt, a factor but their capacity to contribute to amplifying our understanding of physical reality and, even, of new philosophical ideas, was also stressed as an important element (Ortiz, 2012).

It cannot be said that Einstein's visit determined a clear and sustained transition in Argentina's scientific activities in the direction of topics specifically related to relativity theory; his stay was too short for that. In addition, this possibility was never made explicit or declared to be an objective of his visit or a likely consequence in the large process that led to the physicist arrival in the shores of Argentina. Nor had a policy of foreign scholarships in areas of relativity theory, to support such objective, been part of the discussion, either before or immediately after it.

It should be born in mind that Loedel Palumbo, the most original local relativist at the time, had been introduced to relativity theory by his teacher Gans, in La Plata, *before* Einstein's visit. Later, the young physicist had graduate students and influenced others, but towards the end of the 1920s his own research work moved in a different direction: away from relativity proper and towards the analysis of the foundations of physics.

The general shift in physics research in the direction of the new quantum mechanics, which became dominant in the international scene in the 1930s, clearly diminished the strength of efforts on relativity theory.

However, even taking all those elements into account, there is little doubt that the presence of Einstein in Argentina gave physics a more prominent position in science and provided young physicists with a stronger motivation, especially those who had a chance to benefit personally from the efforts made by the visitor to communicate and share with them his vision on the future of science, whether they were theoretical physicists, mathematicians or philosophers of science.

The visit had also an impact on the topics discussed in some of the more advanced levels of university training in mathematics. The case of Rey Pastor is paradigmatic: this eminent mathematician often included topics related to the mathematical foundations of relativity theory in his advanced courses and seminars. For example, on modern differential geometry in 1923 and 1924, shortly before Einstein's arrival (Rey Pastor, 1989); later, in the 1930s, when a new generation of theoretical physicists was being trained in Argentina (Ortiz, 2012), Rey Pastor returned to topics related to relativity theory; for example, in the mid 1930s, to tensor algebra and special relativity theory, and later to tensor calculus and general relativity. Early in the same decade, Butty introduced a systematic study of tensor calculus in his Mathematical Physics courses at the University of Buenos Aires, and used it to discuss Einstein's theory (Butty, 1931, 1934).

No doubt, in the list of scientific visitors Argentina received in the 20th century, Einstein occupies a position of a considerable interest and importance. This is because of the influence his visit had on the development and on the public perception of mathematics and physics, and also on account of its impact on other areas of culture, particularly on modern philosophy, independent of that impact having been faithful, or not, to the meaning of Einstein's theory of relativity.